\documentclass{aa}  
\usepackage{graphicx}
\usepackage{txfonts}
\usepackage{color}
\usepackage[normalem]{ulem}

\begin{document} 
   \title{Discovery and Follow-up of the Unusual Nuclear Transient OGLE17aaj}


   \author{M.~Gromadzki\inst{1}
          \and
          A.~Hamanowicz\inst{1,2}
          \and
          L.~Wyrzykowski\inst{1}
          \and
          K.~V.~Sokolovsky\inst{3,4,5} 
          \and
          M.~Fraser\inst{6}
          \and
          Sz.~Koz{\l}owski\inst{1}
          \and
          J.~Guillochon\inst{7}
          \and
          I.~Arcavi\inst{8,9,10,11}
          \and
          B.~Trakhtenbrot\inst{10,12,13}
          \and
          P.~G.~Jonker\inst{14,15}
          \and
          S.~Mattila\inst{16}
          \and
          A.~Udalski\inst{1}
          \and
          M.~K.~Szyma\'nski\inst{1}
          \and          
          I.~Soszy\'nski\inst{1}
          \and
          R.~Poleski\inst{1,17}
          \and 
          P.~Pietrukowicz\inst{1}
          \and 
          J.~Skowron\inst{1}
          \and
          P.~Mr\'oz\inst{1}
          \and
          K.~Ulaczyk\inst{1,18}
          \and 
          M.~Pawlak\inst{1}
          \and
          K.~A.~Rybicki\inst{1}
          \and
          J.~Sollerman\inst{19}
          \and
          F.~Taddia\inst{19}
          \and
          Z.~Kostrzewa-Rutkowska\inst{14,15}
          \and
          F.~Onori\inst{14,15}
          \and
          D.~R.~Young\inst{20}
          \and
          K.~Maguire\inst{20} 
          \and
          S.~J.~Smartt\inst{20}
          \and
          C.~Inserra\inst{21}
          \and
          A.~Gal-Yam\inst{22}
          \and
          A.~Rau\inst{23}
          \and
          T.-W.~Chen\inst{23}
          \and
          C.~R.Angus\inst{21}
          \and
          D.~A.~H.~Buckley\inst{24}
          }

   \institute{Warsaw University Astronomical Observatory, Al. Ujazdowskie 4, 00-478 Warszawa, Poland, 
              \email{marg@astrouw.edu.pl}
         \and
         European Southern Observatory, Karl Schwarzschild Str 2, D-85748 Garching, Germany
         \and
         IAASARS, National Observatory of Athens, Vas.~Pavlou \& I.~Metaxa, Penteli~15236, Greece
         \and
         Sternberg Astronomical Institute, Moscow State University, Universitetsky pr. 13, Moscow 119991, Russia 
         \and
         Lebedev Physical Institute, Astro Space Center, Profsoyuznaya 84/32, Moscow 117997, Russia
         \and
         School of Physics, O'Brien Centre for Science North, University College Dublin, Belfield, Dublin 4, Ireland
         \and
         Harvard-Smithsonian Center for Astrophysics, 60 Garden St, Cambridge, MA 02138, USA
         \and
         Department of Physics, University of California, Santa Barbara, CA 93106-9530, USA
         \and
         Las Cumbres Observatory, 6740 Cortona Dr Ste 102, Goleta, CA 93117-5575, USA
         \and
         The Raymond and Beverly Sackler School of Physics and Astronomy, Tel Aviv University, Tel Aviv 69978, Israel. 
         \and
         Einstein Fellow
         \and
         Department of Physics, ETH Zurich, Wolgang-Pauli-Str. 27, Zurich, CH-8093, Switzerland
         \and
         Zwicky Fellow
         \and
         SRON, Netherlands Institute for Space Research, Sorbonnelaan 2, NL-3584 CA Utrecht, the Netherlands
         \and
         Department of Astrophysics/IMAPP, Radboud University, PO Box 9010, NL-6500 GL Nijmegen, the Netherlands
         \and
         Tuorla Observatory, Department of Physics and Astronomy, University of Turku, V\"ais\"al\"antie 20, FI-21500 Piikkia\"o, Finland
         \and
         Ohio State University, Department of Astronomy, 140 West 18th Avenue, Columbus, OH, USA 43210
         \and
         Department of Physics, University of Warwick, Gibbet Hill Road, Coventry, CV4 7AL, UK
         \and
         Department of Astronomy, The Oskar Klein Center, Stockholm University, AlbaNova, 10691, Stockholm, Sweden
         \and
         Astrophysics Research Centre, School of Mathematics and Physics, Queens University Belfast, Belfast BT7 1NN, UK
         \and
         School of Physics \& Astronomy, University of Southampton, Southampton, SO17 1BJ, UK
         \and
         Department of Particle Physics and Astrophysics, Weizmann Institute of Science, 234 Herzl St., Rehovot, Israel
         \and
         Max-Planck-Institut für Extraterrestrische Physik, Giessenbachstr. 1, 85748 Garching, Germany
         \and
         South African Astronomical Observatory, PO Box 9, Observatory 7935, South Africa
         }

   \date{Received ; accepted }

 
  \abstract
   {}
   {We report on the discovery and follow-up of a peculiar transient, OGLE17aaj, which occurred in the nucleus of a weakly active galaxy. We investigate whether it can be interpreted as a new candidate for a tidal disruption event (TDE).}
   {We present the OGLE-IV light curve that covers the slow 60-day-long rise to maximum along with photometric, spectroscopic, and X-ray follow-up during the first year.}
   {OGLE17aaj is a nuclear transient exhibiting some properties similar to previously found TDEs, including a long rise time, lack of colour-temperature evolution, and high black-body temperature. On the other hand, its narrow emission lines and slow post-peak evolution are different from previously observed TDEs. Its spectrum and light-curve evolution is similar to F01004-2237 and AT~2017bgt. Signatures of historical low-level nuclear variability suggest that OGLE17aaj may instead be related to a new type of accretion event in active super-massive black holes.}
   {}

   \maketitle
%

\section{Introduction} 
Transients in the centres of galaxies are challenging to discover and study. However, an increasing number of such flares is being detected by wide-field sky surveys, such as iPTF \citep{Blagorodnova2017}, Pan-STARRS \citep{Lawrence2016}, and OGLE \citep{Wyrzykowski2017}.
A variety of flaring mechanisms have been suggested to explain these discoveries, which range from supernovae, changing-look active galactic nuclei \citep[AGN; e.g.][]{LaMassa2015}, extreme AGN flares of unclear nature \citep[e.g.][]{Graham2017}, or tidal disruption events (TDEs). A TDE occurs when 
a star passing near a super-massive black hole (SMBH)
residing in the centre of a galaxy is disrupted by tidal forces.
Roughly half of the material of the disrupted star forms an accretion flow around the SMBH \citep{Rees1988}. 
The interaction of the matter in the accretion flow produces a bright, blue flare that lasts from months to years. 
For SMBHs with mass $\lesssim 10^6 M_{\odot}$, the flare amplitude is expected to be an order of magnitude lower  than for a $\gtrsim 10^7 M_{\odot}$ SMBH, resulting in lower peak luminosities and posing additional challenges for detection \citep{Guillochon2015}.
The absolute magnitude of a TDE in the optical can be very bright and may reach $M = -20$ or even up to $M = -23$ \citep{Arcavi2014,Leloudas2016}. However, the recent discovery of the fast and faint event iPTF16fnl~\citep{Blagorodnova2017} suggests that the population might also extend to fainter peak magnitudes. 

The estimated TDE rate is about one per $10^4 - 10^5$ years per galaxy \citep{Wang2016,Stone2016}. However, most optically selected TDEs have been discovered in post-starburst (E+A) galaxies \citep{Arcavi2014}, which are elliptical galaxies with an atypically large population of young A-type stars. It is not clear whether these rates also apply to other types of hosts, as some recent TDE candidates have been found in different types of galaxies: in weak AGN \citep[OGLE16aaa;][]{Wyrzykowski2017} or ultra-luminous infrared galaxies \cite[ULIRGs;][]{Tadhunter2017,Mattila2018}. These discoveries show that our current knowledge of TDE demographics is far from complete. 

We here report the discovery and early follow-up of a nuclear transient, OGLE17aaj, which shares spectral features with F01004-2237 \citep{Tadhunter2017}. Recently, another transient (AT~2017bgt) exhibited similar properties, and its TDE interpretation has been strongly disfavoured \citep{Trakhtenbrot}. Understanding the nature of OGLE17aaj can therefore shed light on this type of phenomena.
In Sect. 2 we describe the observations, and we present the results in Sect. 3. We discuss the results in Sect. 4 by comparing OGLE17aaj to other known TDEs. We summarise in Sect. 5.
We adopt the following cosmological parameters: $H_0$ = 70 km $\textrm{s}^{-1}$ $\textrm{Mpc}^{-1}$, $\Omega_m$ = 0.28, and $\Omega_{\Lambda}$ = 0.72. 

\section{Observations} 
OGLE17aaj is located at $\alpha_{\rm J2000}$ = 01:56:24.93, $\delta_{\rm J2000}$ = $-$71:04:15.7. We measured this position to be 0.10$\pm$0\farcs13 (equivalent to 0.21$\pm$0.27 kpc at z=0.116) from the centre of its host galaxy, GALEXASC J015624.70-710415.8, and it is therefore coincident with the host nucleus. 

The transient was discovered on the night of 2017 January 2 \citep{Gromadzki2017} by the OGLE-IV Transient Detection System \citep{Wyrzykowski2014,Klencki2016}. 
The field within which OGLE17aaj was found has been systematically monitored since 2010 by the OGLE-IV project, with cadences of 5-10 days and two months in $I$ and $V$ band, respectively. The limiting magnitudes in both bands are about $22^{m}$ (Vega).  
Difference-imaging photometry at the position of OGLE17aaj was extracted using a stack of the best-seeing images taken prior to the flare as a deep reference image. Details of OGLE-IV can be found in \citet{Udalski2015}.
The earliest detection of the event dates from 2016 October 30, and the event reached its $I$-band maximum around 2017 January 4. OGLE collected 112 and 9 observations in $I$ and $V$ band, respectively.

OGLE17aaj was observed using the 8.2m Antu telescope of the Very Large Telescope (VLT) with FORS2 in long-slit mode. Observations were taken on the nights of 2017 January 15, February 2, and 27 and October 6, under seeing conditions of 0.4-0\farcs7, and at air masses 1.7-2.0. The first and last spectra were taken with the G300V grism, and the remaing were taken with G600B. In all cases the exposure time was 974 s. 
These spectra were bias-corrected, flatfielded, extracted, and wavelength and flux calibrated using calibration frames provided by ESO and standard IRAF procedures.

The transient was also observed using the 3.5m New Technology Telescope at La Silla, using the EFOSC2 spectrograph with Gr11. Observations were taken on the nights of 2017 January 20 and 26 and February 19 under good conditions, with seeing of 0.7-0\farcs9, and at an air mass of $\sim$1.7, as part of the PESSTO ESO public survey \citep{Smartt2015}. The exposure time for each spectrum was 2700 s. All EFOSC2 spectra were reduced using the PESSTO pipeline. 

An additional spectrum was obtained with the Southern African Large Telescope (SALT) using the Robert Stobie Spectrograph (RSS), equipped with the PG300 grating. Two exposures of 1200 s were taken on 2017 August 18, under good conditions with a seeing of 1\farcs0 and an air mass $\sim$1.3. Basic reductions were performed using the PySALT package \citep{Crawford2010}, and IRAF was used for extraction and wavelength and flux calibration.   
 
The \textit{Neil Gehrels Swift Observatory} \citep[\textit{Swift};][]{Gehrels2004} observed OGLE17aaj for 14 epochs (19.1\,ks total exposure time) between 2017 January 24 and September 9, with both the X-Ray Telescope (XRT; \citealt{Burrows2005}; operating in the photon counting mode) and the Ultra-Violet/Optical Telescope (UVOT; \citealt{Roming2005}).
The UVOT calibration of \citet{Poole2008} and \citet{Breeveld2010} was used to convert UVOT count rates into magnitudes and flux densities. We used archival {\it GALEX} flux measurements for host subtraction.


\section{Results}
\begin{figure}
\includegraphics[width=0.48\textwidth,clip=true,trim=0.8cm 0.5cm 1.2cm 1.0cm]{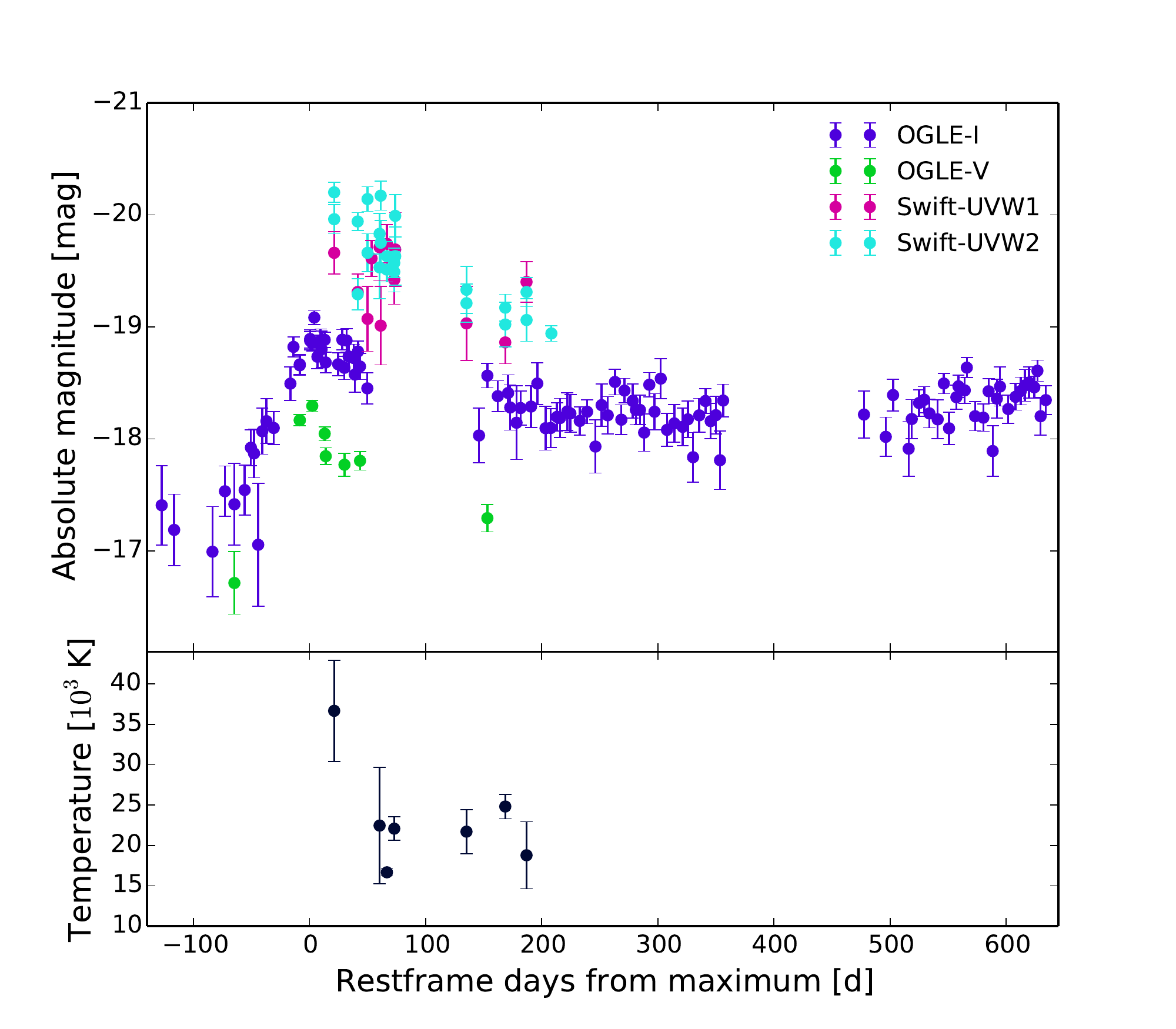}
\caption{Upper panel: Host-subtracted   photometry of OGLE17aaj in the OGLE-IV $I$ and $V$ bands and the {\it Swift} UVW1 and UVW2 bands in absolute magnitudes as a function of days since maximum. Lower panel: Temperature evolution derived from {\it Swift} ultraviolet data over the same period.}
\label{fig:multicolorlightcurve}
\end{figure}

OGLE17aaj was detected at a host-subtracted $I$-band magnitude of about 21\fm0 and  reached an $I$-band maximum of 20$^{\rm m}$ in about 60 days (Fig.\,\ref{fig:multicolorlightcurve}). The transient then declined for about 200 days and appears to subsequently have reached a plateau. The UV measurements follow the trend seen in optical observations. At maximum, the transient reached $m_{UVW1}=19\fm06$ and $m_{UVW2} = 18\fm52$ on galaxy-subtracted {\it Swift} images, which corresponds to an increase in UV flux by a factor of $\sim$16 compared to archival {\it GALEX} measurements. The estimated peak absolute magnitude of OGLE17aaj in the $I$ band is $M_I = -18\fm8$. Because the transient is not yet completed, we can derive the lower limit on the total energy emitted until now as about 7$\times$$10^{51}$ ergs, which corresponds to a lower limit of the accreted mass of $M_{acc}$$\sim$0.04 $M_{\odot}$, assuming a radiating efficiency of 0.1.

The UV source associated with the host galaxy is listed in the {\it GALEX} catalogue \citep{Bianchi2011} as GALEXASC\,J015624.70$-$710415.8, with far-UV (FUV) AB magnitudes of $21\fm414\pm0\fm305$ and near-UV (NUV) $21\fm183\pm0\fm172$. 
No X-ray sources are detected in the {\it ROSAT} all-sky survey \citep{Boller2016} in the vicinity of this galaxy. No radio catalogues comparable in depth to the ATCA post-outburst observations \citep[non-detection at the 50\,$\mu$Jy level at 5 and 9\,GHz;][]{Stanway2017} cover this region of the sky. The deepest surveys are the AT20G survey, conducted at 20\,GHz with ATCA \citep[flux-density limit of 40\,mJy;][]{Murphy2010} and the GLEAM survey at 72--231\,MHz with the Murchison Widefield Array \citep[50\,mJy;][]{Hurley-Walker2017}. The host was measured by the Wide-Field Infrared Survey Explorer, $W1$=15\fm113$\pm$0\fm030, $W2$=14\fm889$\pm$0\fm046, $W2$=14\fm889$\pm$0\fm046, $W3$=12\fm089$\pm$0\fm223, and $W4$>8\fm953 \citep[AllWISE;][]{Wright2010,Cutri2013}. Photometric data for this galaxy can also be found in the USNO-B1.0 catalogue \citep{Monet2003}.
We used these archival data, that is, broad-band magnitudes from GALEX, USNO-B1.0, and AllWISE,  
to estimate the spectral energy density (SED) of the host before 
the appearance of the transient \citep[following][]{Kozlowski2015}. To fit the SED, we used the low-resolution templates of AGN and galaxies from \citet{Assef2010}.  
The best-matching template to explain the observed SED comprised an elliptical galaxy plus an AGN, but the residuals of the host light S\'ersic model fit with GALFIT \citep{Peng2010} indicate weak spiral arms.

\begin{figure}
\includegraphics[width=0.48\textwidth,clip=true,trim=0.0cm 0.0cm 0.0cm 0.0cm]{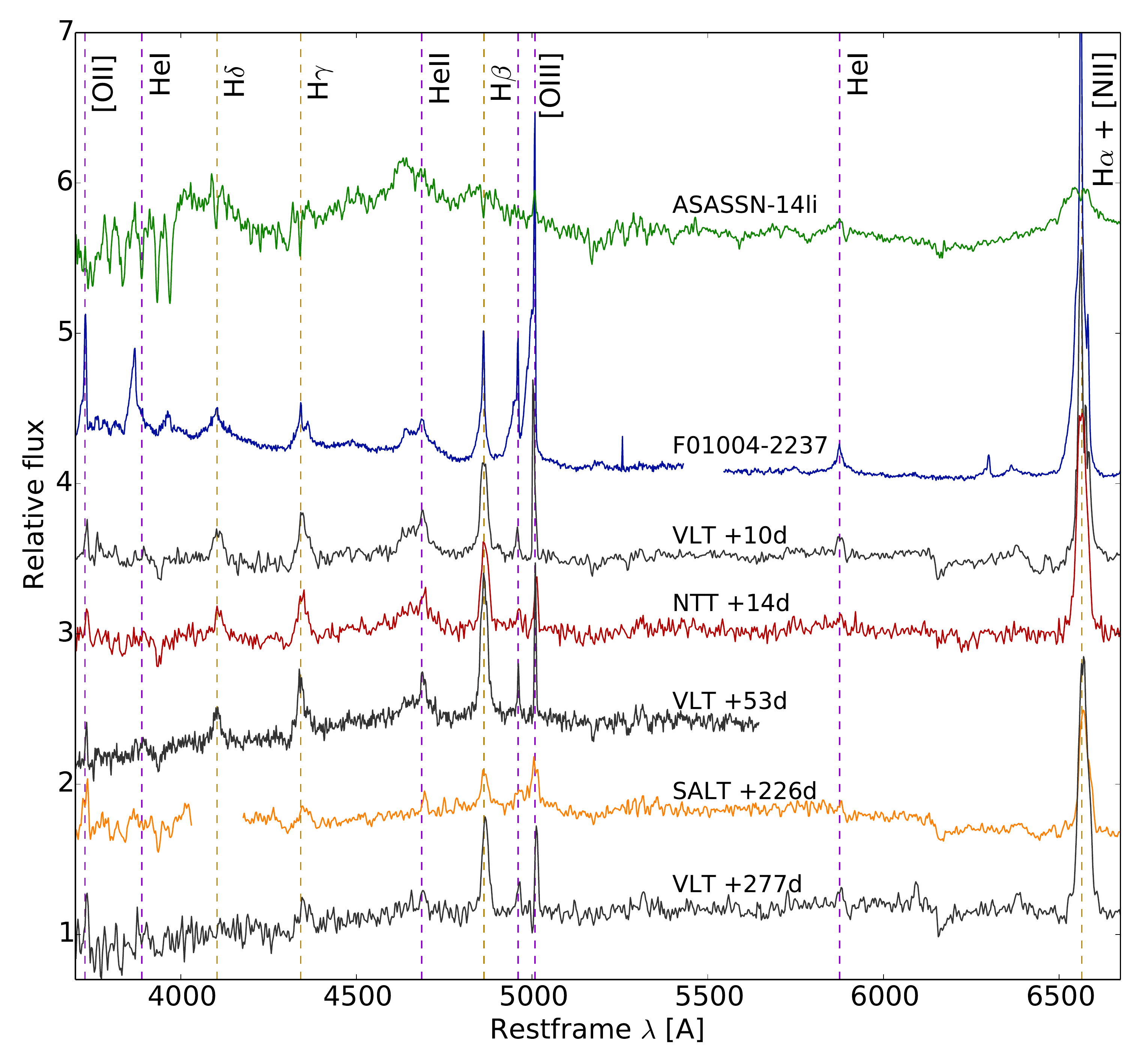}
\caption{Sequence of spectra obtained for OGLE17aaj. VLT observations are marked in black, those from the NTT in red, and SALT observations are shown in orange. The upper two spectra are of F01004-2237 \citep{Tadhunter2017} and a typical TDE, ASASSN-14li \citep{Holoien2016}, for comparison.}
\label{fig:spectraltimeseries}
\end{figure}

The classification spectrum, taken ten days after maximum (Fig.\,\ref{fig:spectraltimeseries}), revealed a flat continuum with many prominent emission lines: hydrogen Balmer lines, \ion{He}{I} $\lambda$5876, \ion{He}{II} $\lambda$4686, [\ion{O}{III}] $\lambda\lambda$4960, 5007, [\ion{O}{II}] $\lambda$3727, and [\ion{S}{II}] $\lambda\lambda$6718, 6732. 
The lines are at a redshift of $z=0.116$, which corresponds to a luminosity distance of 540 Mpc .
The spectrum looks very similar to the spectra of F01004-2237 from September 2015, that is, to spectra{\it } taken around five years after its maximum, when the transient was still ongoing.
The shape of the line complex around \ion{He}{ii}~$\lambda$4686 in particular is very similar (see Fig. \ref{fig:spectraltimeseries}).
It has a double-peaked structure, where the bluer broad component can be associated with \ion{N}{iii} $\lambda$4640, or blue-shifted \ion{He}{ii}. The FWHM of the combined double feature, $\sim$5000 km s$^{-1}$, is also very similar in both objects.
The line ratio of (\ion{N}{iii}+\ion{He}{ii})/H$\beta$$\approx$0.9 is twice lower than in  F01004-2237, while typical quasars have $\sim$0.02 \citep{vandenBerk2001}.

During the first two months of spectroscopic follow-up, the spectrum of OGLE17aaj essentially remained stable. The equivalent widths of the emission lines varied at a level of 20 percent. However, it is unclear whether these are caused by real, astrophysical effects or by the varying quality of the spectra, which were usually taken at high air masses (1.7-2). After a seasonal gap, we took another SALT spectrum, in which the double-peak profile around \ion{He}{ii} disappeared and only one emission feature remained. This spectrum also shows a decrease in the flux of the H$\beta$ and H$\alpha$ emission lines (by a factor of 3 and 1.5, respectively). We are unable to compare the SALT +226d spectrum with previous VLT spectra in detail because SALT used a slit that is twice as wide, and therefore the contribution from host emission was more significant.
We took an additional VLT spectrum with the same configuration as the first, which shows that the intensities of \ion{He}{ii} and H$\gamma$ decrease by a factor of $\sim$3, and H$\delta$ practically disappeared. 
We also attempted to estimate the position of the host on the BPT diagnostic diagram, but because our spectra contain both the host and transient, the line measurements were strongly affected by the transient, which increased the flux in hydrogen lines. Such measurement will be possible after the transient is completed.

The shape of the continuum is rather flat, although Swift colours suggest a high temperature of $\sim$20\,000 K, and it recalls those in Seyfert type II galaxies. The main difference is the presence of \ion{N}{III}$+$\ion{He}{II} emission, and the Balmer emissions are also much stronger. Most likely, an underlying, persistent AGN continuum contributes to the optical spectrum, and the difference imaging therefore merely shows that the variable component of this flat continuum is blue. A behaviour like this is also seen in variable AGN \citep[e.g.][]{Hung2016}. 

A weak X-ray source is detected at the $4\sigma$ confidence level in the stacked {\it Swift}/XRT image at the position of the transient, with a net count rate of 0.0011$\pm$0.0003\,cts/s. When we fix the \ion{H}{i} column density to the Galactic value along this line of sight
($5.53 \times 10^{20}$\,cm$^{-2}$; \citealt{Bajaja2005,Kalberla2005}), the spectrum can be fitted by an absorbed power law with a photon 
index of $2.5 \pm0.6$ and an intrinsic 0.3--10\,keV flux of $4.5 \times 10^{-14}$\,ergs cm$^{-2}$ s$^{-1}$. 
The corresponding intrinsic luminosity in this band is $1.57 \times 10^{42}$\,ergs s$^{-1}$. The flux estimation uncertainty is about 40\%. 
The [\ion{O}{III}] $\lambda$5007 to hard X-ray ratio is 
consistent with what is typically seen in X-ray detected, low-redshift AGN  \citep[e.g.][and references therein]{Berney2015}. 

No detectable X-ray emission is found in individual observations with a typical upper limit of 0.002\,cts/s ($8 \times 10^{-14}$\,ergs cm$^{-2}$ s$^{-1}$).
Taking into account the Galactic reddening of $E(B-V)=0\fm025$ \citep{Schlafly2011}, we estimated the colour temperature (Fig.\,\ref{fig:multicolorlightcurve}, bottom panel) using the three {\it Swift}/UVOT ultraviolet filters, where the host light contribution is minimal. 
We estimated the mass of the SMBH using the properties of the host galaxy based on the pre-transient imaging. 
The fraction of the light from the bulge with respect to the total flux emitted by the galaxy, measured with GALFIT on a deep stack of $V$-band archival OGLE images, is 19.8\%, which yields $\log M_{BH}$=7.37 using the method of \citet{Bentz2009}.

\begin{figure}
\includegraphics[width=0.48\textwidth]{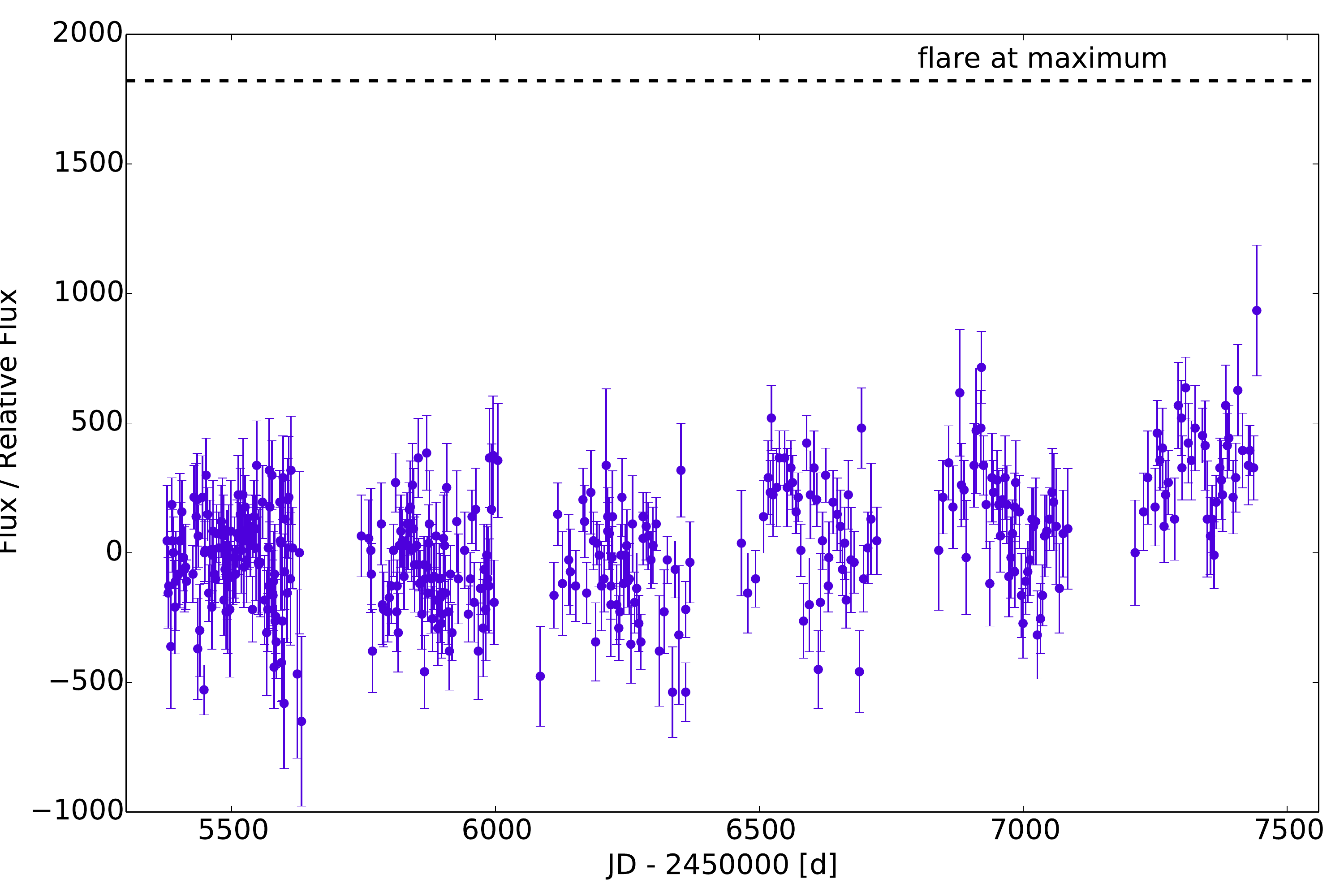}
\caption{OGLE $I$-band host-subtracted flux measurement at the centre of the host galaxy before the transient, covering 2010-2016. The horizontal line shows the maximum flux level reached by OGLE17aaj in 2017.}
\label{fig:archival}
\end{figure}

Additionally, the archival OGLE-IV $I$-band light curve, covering six years before the transient, shows a small-scale irregular variability at a level of 10\%, which is significantly lower than the flare itself (Fig.\,\ref{fig:archival}) and indicates on-going accretion onto the SMBH. 
\section{Discussion}
The UV and optical light curves of OGLE17aaj indicate no or slow colour-temperature change over the entire monitoring period. Temperatures as high as $4 \times 10^4$\,K at maximum are not typical of supernovae, but are seen in TDEs \citep[e.g.][]{Gezari2012,Arcavi2014}.
The peak absolute magnitude of OGLE17aaj 
($M_{I}=-18.8$) is fainter than in most known optical TDEs and exceeds $-20^m$. Nevertheless, it is not the first example of such a faint event, as the faint TDE iPTF16fnl reached only $M_g = -17.2$ \citep{Blagorodnova2017}. 

The light curve from OGLE-IV shows a well-covered, smooth, and long-duration rise to maximum. Most of the TDEs reported so far hardly ever had such a good pre-discovery coverage. While the slow rise of OGLE17aaj could still be in agreement with an interpretation as a TDE, the decline is much slower (about 0.5 mag/year). In particular, the flattening of the light curve after 200 days has never been seen in any TDE. The light curve of OGLE17aaj is somewhat similar to the changing-look AGN SDSS J233317 \citep{MacLeod2016}, but the spectrum of that object does not have the characteristic \ion{He}{II}$+$\ion{N}{III} feature seen in OGLE17aaj.
This suggests that OGLE17aaj may be related to a change in the state of the AGN rather than a TDE. 

A broad ($\sim$30\,000 km s$^{-1}$) \ion{He}{II} line component is the most typical feature of most TDEs (e.g. \citealt{Arcavi2014}). 
However, in the case of OGLE17aaj, this emission line is significantly narrower at $\sim$5\,000 km s$^{-1}$.
The \ion{He}{II} profile is irregular and is very similar to what is seen in the F01004-2237 event five years after its maximum \citep{Tadhunter2017}.
The complex feature around \ion{He}{II} ~$\lambda$4686 in F01004-2237 was explained with a large population of $\sim10^4$ Wolf-Rayet stars (WR), and with an additional blue-shifted helium line with an FWHM$\sim$6200 km s$^{-1}$, which, according to \citet{Tadhunter2017}, originated from a TDE.

The host of OGLE17aaj shows weak levels of AGN activity. In contrast to F01004-2237, it has a negligible WR population.  However, the similar optical spectra and the slow photometric evolution suggest that they may be driven by a similar phenomenon.  We note that the estimated black hole mass for the SMBH we associate with OGLE17aaj is an order of magnitude lower than in the case of F01004-2237. 
Both the archival photometric variability of the nucleus and narrow lines in the spectrum originating from the host indicate that the nucleus of the galaxy is active at a low level. If the SMBH has been accreting in the past, the observed greater change in brightness fits the overall picture of a rapid change in the accretion rate.

Recently, \citet{Trakhtenbrot} have reported the discovery of the transient AT~2017bgt, which showed optical features that are very similar to those seen in OGLE17aaj, as well as a similar photometric evolution. 
High-quality near-IR and optical spectra of AT~2017bgt showed single-peaked \ion{He}{ii} lines, which rules out a disk origin for the double-peak profile in the broad feature associated with \ion{He}{II} $\lambda$4686, and thus suggests that the bluer peak originates from \ion{N}{iii} $\lambda$4640. They also identified several blue \ion{O}{iii} lines and argued that the \ion{O}{iii} and \ion{N}{iii} features are produced by the Bowen fluorescence mechanism, which is driven by enhanced accretion of gas onto an existing AGN. This further disfavours a TDE.

\section{Conclusions}
We reported the discovery and early follow-up observations of the unusual nuclear transient OGLE17aaj. 
We investigated it as a potential TDE candidate based on its high black-body temperature and on the similarity of its spectrum to that of the recently reported nuclear event in F01004-2237, which was interpreted as a TDE \citep{Tadhunter2017}. 
The narrow spectral features of the host and pre-flare photometric variability might instead indicate that these types of events represent a change in the accretion flow of AGN.
OGLE17aaj showed a well-covered 60-day-long rise and a slow 200-day decline, followed by a plateau until day 300. This is 
unlike any TDE reported so far.

The discovery of OGLE17aaj, as well as other similar transients with high amplitudes in AGN, indicates that a new class of AGN-related phenomena exist. It may lead to new ways of investigating the close environments of SMBHs and the accretion flows through which they grow.
\begin{acknowledgements}
OGLE project is supported by the Polish NCN MAESTRO grant 2014/14/A/ST9/00121 to AU.
This work is based on ESO programmes: 098.B-0768 (PI: Wyrzykowski), 0100.B-0503 (PI: Gromadzki), 188.D-3003, 191.D-0935 (PI: Smartt) and SALT Programme 2016-2-LSP-001. Polish participation in SALT is funded by grant no. MNiSW DIR/WK/2016/07.
MG, AH, LW and KAR were supported by the Polish National Science Centre grant OPUS 2015/17/B/ST9/03167.
PGJ, ZKR and FO acknowledge support from ERC Consolidator Grant 647208.
MF is supported by a Royal Society - Science Foundation Ireland University Research Fellowship.
KM acknowledges support from the STFC through an Ernest Rutherford Fellowship.
Support for IA was provided by NASA through the Einstein Fellowship Program, grant PF6-170148.
DB acknowledges support from the National Research Foundation of South Africa. 
\end{acknowledgements}
%
%

\end{document}